\documentclass[useAMS,usenatbib]{mn2e}

\usepackage{graphicx}
\title[Reduction of the GLE to a 1D non-linear form]
{Reduction of the gravitational lens equation to a one-dimensional 
non-linear form for the tilted Plummer model family}

\author[Francisco Frutos-Alfaro]{Francisco Frutos-Alfaro
\thanks{Present address: Theoretical Astrophysics Institute, 
University of T\"ubingen, Auf der Morgenstelle 10C, 72076 T\"ubingen, 
Germany, Email: frutto@tat.physik.uni-tuebingen.de} \\
{Space Research Centre and School of Physics} \\ 
{University of Costa Rica} \\
{San Jos\'e, Costa Rica}}

\begin{document}

\date{Accepted 2007 January 11. Received 2006 December 29; 
in original form 2006 October 23}

\pagerange{\pageref{firstpage}--\pageref{lastpage}} 
\pubyear{2007}

\maketitle

\label{firstpage}

\begin{abstract}
The gravitational lens equation for the tilted Plummer family of models can be 
reduced to a one-dimensional non-linear equation. For certain values of the 
slope of the radial profile it can be reduced to a polynomial form. Both forms 
are advantageous to find the roots, i.e. the images for a given model. 
The critical curve equations can also be reduced to a non-linear or polynomial 
form, and therefore it is useful to find the caustics. This lens model family 
has ample use in gravitational lens theory, and can produce up to five images.
\end{abstract}


\section{Introduction}

\noindent
Smooth non-singular isothermal sphere (SNIS) models are often used in 
gravitational lens theory. Including ellipticity, these models are called 
elliptical SNIS (ESNIS) \citep{blanko} or tilted Plummer family 
\citep{kaskov}. This class of models can be generated from elliptical mass 
distributions or elliptical potentials. The models obtained by means of mass 
distributions are realistic even for higher ellipticities. The models 
generated using the elliptical potentials are restricted to small 
ellipticities \citep{kaskov}, because the isodensity contours become 
dumbbell-shaped for higher ellipticities. Moreover, the density for these 
elliptical potencials could be negative for certain combinations of parameters 
\citep{kaskov}. For more information about these models, the interested reader 
may consult the references at the end of this Letter.

\noindent
For ESNIS models, the gravitational lens equation (GLE) becomes non-linear, 
and hard to solve if the external perturbations (density and shear) are taken 
into account. However, for ESNIS models obtained from elliptical potentials, 
it is possible to simplify this equation to a one-dimensional non-linear or 
polynomial form. We will show how to do it.

\noindent
The advantage of reducing the GLE to a one-dimensional non-linear or 
polynomial form is that there is simple iterative methods to get the roots of 
this class of equations. The critical curve equation can also be 
reduced to a non-linear or polynomial form. This allows us to find 
the caustics easily. Examples of these models will be presented in this Letter.

\section[]{ESNIS Models}

\noindent
The tilted Plummer family of elliptic potentials has the 
form% \citep{sef}

\begin{equation} 
\label{potential}
\psi ({\bf x}) = \left\{
\begin{array}{cc}
\frac{\kappa_0}{2 \beta} 
(x^2_0 + \varepsilon_1 x^2_1 + \varepsilon_2 x^2_2)^{\beta} - \psi_0 , 
& 0 < \beta < 1 \\
\frac{\kappa_0}{2} \ln{(x^2_0 + \varepsilon_1 x^2_1 + \varepsilon_2 x^2_2)} , 
& \beta = 0 ,
\end{array}
\right.
\end{equation} 

\noindent
where $ {\bf x} = (x_1, \, x_2) $, $ x_1 $ and $ x_2 $ are measured in the 
direction of the principal axes of the ellipsoid (image position), $ \beta $ 
is the slope of the radial profile of the potential, $ x_0 $ is the core 
radius (same units as $ x_1 $ and $ x_2 $), $ \kappa_0 $ is a constant, 
$ \varepsilon_i = 1 \pm \varepsilon $ ($ i = 1, \, 2 $) with $ \varepsilon $ 
as the lens ellipticity, and $ \psi_0 $ is a constant potential reference, 
which will be taken as null. 

\noindent
The scaled deflection angle components for the ESNIS models have 
the following form:

\begin{equation} 
\label{alpha}
\alpha_i ({\bf x}) = 
\frac{\kappa_i x_i}{(x^2_0 
+ \varepsilon_1 x^2_1 + \varepsilon_2 x^2_2)^{1 - \beta}} ,
\end{equation} 

\noindent
where $ \kappa_i = \kappa_0 \varepsilon_i $.

\noindent
The surface mass density is given by 

\[ 
\kappa ({\bf x}) = \kappa_0 
(x^2_0 + \varepsilon_1 x^2_1 + \varepsilon_2 x^2_2)^{\beta - 2} [x_0^2 
+ \beta_1 \varepsilon_1 x^2_1 + \beta_2 \varepsilon_2 x^2_2] ,
\]

\noindent
where $ \beta_i = 1 + (\beta - 1) \varepsilon_i $.

\noindent 
To avoid negative values of surface mass density and the dumbbell form of the 
isodensity contour shape, the ellipticity $ \varepsilon $ should be restricted 
to small values \citep{kaskov}: 

%\[
%\varepsilon_{convex} := 
%\frac{\beta}{3 - \beta} < \varepsilon < %\varepsilon_{critic} := 
%\frac{1 - \beta}{\beta} .
%\]

\[
\varepsilon \le \frac{\beta}{3 - \beta} .
\]

\noindent
Among the models, that the ESNIS class contains are:

\begin{itemize}
\item power-law mass ($ \varepsilon = 0, \, \rm{and} \; x_0 = 0 $),
\item NIS ($ \beta = 1 / 2, \, \rm{and} \; \varepsilon \le 1 / 5 $), 
\item Plummer ($ \beta = \varepsilon = 0 $).
\end{itemize}

\section[]{Reduction to a Non-linear or Polynomial Form}

\noindent
The scaled ray tracing equation, or GLE is \citep{sef}

\begin{equation}
\label{gleq} 
{\bf y} = {\sf M} %\cdot 
{\bf x} - {\balpha} ({\bf x}) , 
\end{equation} 

\noindent
where $ {\bf y} $ is the source position on the source plane, and the matrix 
$ {\sf M} $ is given by

\begin{equation}
\label{matrix} 
{\sf M} = \left(
\begin{array}{cc}
1 - \sigma - \gamma \cos{2 \phi} &            - \gamma \sin{2 \phi} \\
           - \gamma \sin{2 \phi} & 1 - \sigma + \gamma \cos{2 \phi}
\end{array}
\right) ,
\end{equation} 

\noindent
where $ \sigma $, and $ \gamma $ are the density and shear of external 
perturbations (nearby galaxies or a cluster contribution), and $ \phi $ is 
the shear angle.

\noindent
Let us define some variables for the sake of simplicity

\begin{eqnarray}
\label{definitions} 
\chi & = & x^2_0 + \varepsilon_1 x^2_1 + \varepsilon_2 x^2_2 , \\
   f(\chi) & = & {\chi}^{1 - \beta} \Rightarrow 
\chi(f) = f^{1 / (1 - \beta)} . \nonumber 
%M_{11} & = & 1 - \sigma - \gamma \cos{2 \phi} \nonumber \\
%M_{12} & = & M_{21} = - \gamma \sin{2 \phi} \nonumber \\
%M_{22} & = &1 - \sigma + \gamma \cos{2 \phi} . \nonumber
\end{eqnarray} 

\noindent
With these definitions the GLE system takes the form

\begin{eqnarray} 
\label{eq1}
y_1 & = & \left(M_{11} - \frac{\kappa_1}{f} \right) x_1 + M_{12} x_2 , \\
\label{eq2}
y_2 & = & \left(M_{22} - \frac{\kappa_2}{f} \right) 
x_2 + M_{21} x_1 .
\end{eqnarray} 

\noindent
Solving for $ x_1 $ and $ x_2 $, we get

%After multiplying (\ref{eq1}) by $ (M_{22} - {\kappa_2} / {f} $, and 
%(\ref{eq2}) by $ M_{12} $, we get by subtracting them

\begin{eqnarray} 
\label{eq3}
g x_1 & = & \left(M_{22} - \frac{\kappa_2}{f} \right) y_1 
- M_{12} y_2 =: f_1 , \\
\label{eq4}
g x_2 & = & \left(M_{11} - \frac{\kappa_1}{f} \right) y_2 
- M_{21} y_1 =: f_2 ,
\end{eqnarray} 

\noindent
where 

\begin{equation} 
\label{det}
g = \left(M_{11} - \frac{\kappa_1}{f} \right) 
\left(M_{22} - \frac{\kappa_2}{f} \right) - M_{12} M_{21} .
\end{equation} 

\noindent
Substituting (\ref{eq3}) and (\ref{eq4}) into (\ref{definitions}) yields 

\begin{equation} 
\label{poly}
g^2 \chi(f) = x_0^2 g^2 + \varepsilon_1 f_1^2 + \varepsilon_2 f_2^2 .
\end{equation} 

\noindent
By substitution of the right-hand side of equations (\ref{eq3}), (\ref{eq4}),
and (\ref{det}) into the last equation, we obtain

\begin{equation} 
\label{polf}
(\chi(f) - x_0^2) P_1(f) = P_2(f) ,
\end{equation} 

\noindent
where

\begin{eqnarray} 
\label{pol1}
P_1(f) & = & [f^2 g]^2 \\ 
& = & [(M_{11} f - {\kappa_1})(M_{22} f - {\kappa_2}) 
- M_{12} M_{21} f^2]^2 \nonumber \\
& = & [{\cal M} f^2 - (\kappa_1 M_{22} - \kappa_2 M_{11}) f 
+ \kappa_1 \kappa_2]^2 , \nonumber 
\end{eqnarray} 

\noindent
and

\begin{eqnarray}
\label{pol2} 
P_2(f) & = & \{\varepsilon_1 [(M_{22} f - {\kappa_2}) y_1 - M_{12} y_2 f]^2 
\nonumber \\
& + & \varepsilon_2 [(M_{11} f - {\kappa_1}) y_2 - M_{21} y_1 f]^2 \} f^2 
\nonumber \\
& = & (A_2 f^2 - A_1 f + A_0) f^2 ,  
\end{eqnarray} 

\noindent
with 

\[
{\cal M} = \rmn{det} [{\bf M}] = M_{11} M_{22} - M_{12} M_{21} ,
\]

\[
A_2 = \varepsilon_1 (M_{22} y_1 - M_{12} y_2)^2 
+ \varepsilon_2 (M_{11} y_2 - M_{21} y_1)^2 , 
\]

\[
A_1 = 2 [\kappa_2 \varepsilon_1 y_1 (M_{22} y_1 - M_{12} y_2) 
+ \kappa_1 \varepsilon_2 y_2 (M_{11} y_2 - M_{21} y_1)]  
\]

\noindent
and

\[
A_0 = \varepsilon_1 (\kappa_2 y_1)^2 + \varepsilon_2 (\kappa_1 y_2)^2 . 
\]

\noindent
$ P_1(f) $ and $ P_2(f) $ are both fourth order polynomials.

\noindent
From (\ref{polf}) we get 

\begin{equation}
\label{polbeta}
f [P_1(f)]^{1 - \beta} = [P_2(f) + x_0^2 P_1(f)]^{1 - \beta} .
\end{equation}

\noindent
The last equation is a non-linear one-dimensional equation for $ f $ that can 
be solved by means of iterative numerical methods, for example, 
the Brent method \citep{brent}. 

\noindent
Now, if we set $ \beta = m / n $, with $ m, \; n$ integers ($ n \ge m $), 
then (\ref{polbeta}) becomes a polynomial form:

\begin{equation}
\label{polmn}
f^{n} [P_1(f)]^{n - m} = [P_2(f) + x_0^2 P_1(f)]^{n - m} .
\end{equation}

\noindent
The image positions can be obtained as follows: first of all solve 
(\ref{polbeta}) or (\ref{polmn}); now substituting the $ f $-values into the 
equations (\ref{eq3}), (\ref{eq4}), and (\ref{det}) yield $ f_1 $, $ f_2 $, 
and $ g$; finally use equations (\ref{eq3}) and (\ref{eq4}) again to get 
$ x_1 $ and $ x_2 $.

%From this polynomial equation, we can read off the number of images or real 
%roots it could have. This model could produce up to $ 5 n - 4 m $ images. 
%Bairstow's method could be applied to solve it. 

\section[]{The Critical Curves and Caustics}

\noindent
The critical curve is the curve formed by all image positions on which the 
determinant of the Jacobian vanishes and the caustic curve is the projection 
of this critical curve to the source plane.

\noindent
The components of the Jacobian 
$ {\sf J} = \partial {\bf y} / \partial {\bf x} $ are

\begin{eqnarray} 
\label{comp}
J_{11} = \frac{\partial y_1}{\partial x_1} & = & 
\left(M_{11} - \frac{\kappa_1}{f} \right) 
+ {2 (1 - \beta) \kappa_1 \varepsilon_1 x_1^2}{\chi^{\eta}} , \nonumber \\
J_{12} = \frac{\partial y_1}{\partial x_2} & = & M_{12} 
+ {2 (1 - \beta) \kappa_1 \varepsilon_2 x_1 x_2}{\chi^{\eta}} , \nonumber \\
J_{21} = \frac{\partial y_2}{\partial x_1} & = & M_{21} 
+ {2 (1 - \beta) \kappa_2 \varepsilon_1 x_1 x_2}{\chi^{\eta}} , \\
J_{22} = \frac{\partial y_2}{\partial x_2} & = & 
\left(M_{22} - \frac{\kappa_2}{f} \right) 
+ {2 (1 - \beta) \kappa_2 \varepsilon_2 x_2^2}{\chi^{\eta}} , \nonumber 
\end{eqnarray} 

\noindent
where $ \eta = \beta - 2 $.

\noindent
The determinant of the Jacobian is

\begin{eqnarray}
\label{detjac}
{\cal J} & = & \rmn{det} [{\sf J}] = J_{11} J_{22} - J_{12} J_{21} \\ 
& = & \frac{1}{[f P_1(f)]^2} \left[
\left((M_{11} f - \kappa_1) P_1(f) 
{\phantom{f^2}} \right. \right. \nonumber \\
& + & \left. \left. 
{2 (1 - \beta) \kappa_1 \varepsilon_1} f^3 [{\cal C}_1 f - \kappa_2 y_1]^2 
\chi^\eta \right) \right. \nonumber \\
& \times & \left. \left((M_{22} f - \kappa_2) P_1(f) 
 {\phantom{f^2}} \right. \right. \nonumber \\
& + & \left. \left. {2 (1 - \beta) \kappa_2 \varepsilon_2} 
f^3 [{\cal C}_2 f - \kappa_1 y_2]^2 \chi^\eta \right) \right. \nonumber \\
& - & \left. \left({2 (1 - \beta) \kappa_1 \varepsilon_2} 
({\cal C}_1 f - \kappa_2 y_1) ({\cal C}_2 f - \kappa_1 y_2) f^3 \chi^\eta 
\right. \right. \nonumber \\
& - & \left. \left. M_{12} f P_1(f) {\phantom{f^2}} 
\! \! \! \! \! \! \right) \right. \nonumber \\
& \times & \left. \left({2 (1 - \beta) \kappa_2 \varepsilon_1}
({\cal C}_1 f - \kappa_2 y_1) ({\cal C}_2 f - \kappa_1 y_2) f^3 \chi^\eta
\right. \right. \nonumber \\
& - & \left. \left. M_{21} f P_1(f) 
{\phantom{f^2}} \! \! \! \! \! \! \right) \right] , \nonumber 
\end{eqnarray}

\noindent
where

\[
{\cal C}_{1} = M_{22} y_1 - M_{12} y_2 
\]

\noindent
and

\[
{\cal C}_{2} = M_{11} y_2 - M_{21} y_1 .
\]

\noindent
In (\ref{detjac}) we have substituted (\ref{eq3}) and (\ref{eq4}). 
From (\ref{detjac}), it is clear that the critical and caustic curves could 
also be found by solving a one-dimensional non-linear equation or a polynomial 
in $ f $.

\section[]{Examples}

\subsection{Plummer's model}

\noindent
Let us take $ \beta = 0 $. We get from (\ref{polbeta}) the following 
polynomial:

\begin{equation}
\label{plummer1}
B_5 f^5 - B_4 f^4 + B_3 f^3 + B_2 f^2 + B_1 f - B_0 = 0 ,
\end{equation}

\noindent
where

\[
B_5 = {\cal M}^2 ,
\]

\[
B_4 = {\cal M}^2 x_0^2 + \varepsilon_1 {\cal C}_1^2 
+ \varepsilon_2 {\cal C}_2^2 + 2 {\cal C}_3 {\cal M} ,
\]

\[
B_3 = 2 \kappa_1 \kappa_2 {\cal M} 
+ (2 x_0^2 {\cal M} + {\cal C}_3) {\cal C}_3 
+ 2 \kappa_2 \varepsilon_1 y_1 {\cal C}_1
+ 2 \kappa_1 \varepsilon_2 y_2 {\cal C}_2 ,
\]

\[
B_2 = 2 \kappa_1 \kappa_2 {\cal C}_3 - \kappa_2^2 \varepsilon_1 y_1^2 
- \kappa_1^2 \varepsilon_2 y_2^2 
- x_0^2 [{\cal C}_3^2 + 2 \kappa_1 \kappa_2 {\cal M}] ,
\]

\[
B_1 = \kappa_1 \kappa_2 [\kappa_1 \kappa_2 - 2 x_0^2 {\cal C}_3] 
\]

\noindent
and

\[
B_0 = [x_0 \kappa_1 \kappa_2]^2 ,
\]

\noindent
with

\[
{\cal C}_{3} = \kappa_1 M_{22} + \kappa_2 M_{11} .
\]

%Source, images, critical curves, and caustics for this model are shown in 
\noindent
The Source, images, critical curves, and caustics for this model are shown in 
Fig. 1. The values of the model parameters are $ \sigma = 0.05 $, 
$ \gamma = 0.2 $, $ \phi = 36^\circ $, $ \varepsilon = 0 $, $ \kappa_0 = 1 $ 
and $ x_0 = 0.5 $. The critical curves and caustics are inclined, because 
we have chosen $ \phi \ne 0 $. 
%In Fig. 2 the isochrones are shown.

%\begin{figure*}%[bhpt]
\begin{figure}%[bhpt]
\begin{center}
\fbox{\includegraphics[width=8.2cm,height=8.2cm]{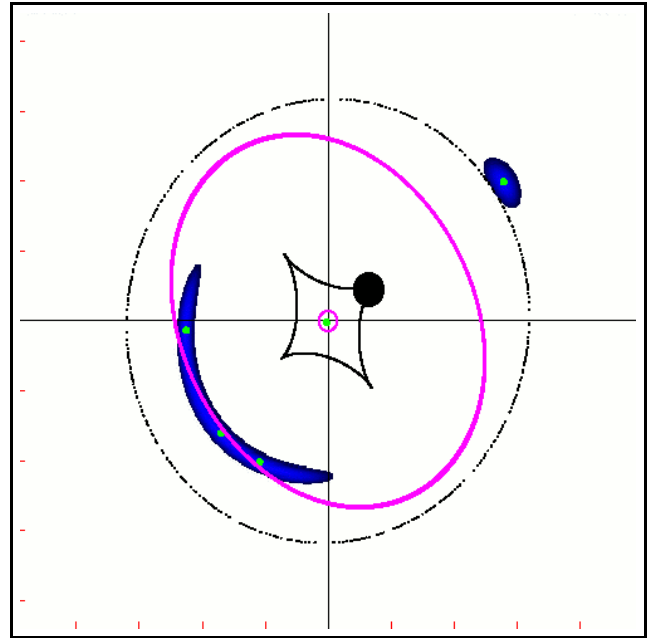}}
\caption{Source (large black dot near the centre), two extended images 
(one of them is a long blue arc) and one tiny image at the centre (green dot), 
the critical curves (magenta ellipse and small circle), and the caustics 
(black diamond and dotted ellipse) produced by the Plummer model for following 
parameter values $ \beta = \varepsilon = 0, \; \gamma = 0.2, \; 
\sigma = 0.05, \; \phi = 36^{\circ}, \kappa_0 = 1, \; x_0 = 1/2 $. 
The green dots are the positions of the images for a point source. 
The position of the extended source is $ (0.65, \; 0.45) $. 
The critical curves and caustics are inclined, because $ \phi \ne 0 $. 
The window size is $ 500 \times 500 $ pixel. 
%The horizontal axis represents 
%$ x_1 $ or $ y_1 $, and the vertical $ x_2 $ or $ y_2 $.
}
\end{center}
\end{figure}
%\end{figure*}

% XFGLenses:
% E = 2.5, theta = 0, alpha = epsilon = 0, gamma = 0.2, sigma = 0.05
% phi = 36, kappa = 1, C = 0.5, X = 0.65, Y = 0.45, R = 0.2

%\begin{figure*}%[bhpt]
%\begin{figure}%[bhpt]
%\begin{center}
%\includegraphics[width=8.4cm,height=8.4cm]{Figure2.ps}
%\caption{Isocrones with images for the Plummer model of Fig 1.
%The dots over the extended images are the images for a point source. 
%In this case there are 5 images: one tiny image close to the centre, 
%three in the arc, and one in the extended image.} 
%\end{center}
%\end{figure}
%\end{figure*}

\subsection{ESNIS model}

\noindent
Let us take $ \beta = 1/2 $. We also get from (\ref{polbeta}) 
a polynomial:

\begin{equation}
\label{plummer2}
B_6 f^6 - B_5 f^5 + B_4 f^4 - B_3 f^3 + B_2 f^2 + B_1 f - B_0 = 0 ,
\end{equation}

\noindent
where

\[
B_6 = {\cal M}^2 ,
\]

\[
B_5 = 2 {\cal C}_3 {\cal M}^2 ,
\]

\[
B_4 = {\cal C}_3^2 + 2 \kappa_1 \kappa_2 {\cal M} 
- [{\cal M}^2 x_0^2 + \varepsilon_1 {\cal C}_1^2 + \varepsilon_2 {\cal C}_2^2],
\]

\[
B_3 = 2 [{\cal C}_3 (x_0^2 {\cal M} - \kappa_1 \kappa_2)  
+ \kappa_2 \varepsilon_1 y_1 {\cal C}_1 
+ \kappa_1 \varepsilon_2 y_2 {\cal C}_2] ,
\]

\[
B_2 = [\kappa_1 \kappa_2]^2 
- x_0^2 [2 \kappa_1 \kappa_2 {\cal M} + {\cal C}_3] 
- \kappa_2^2 \varepsilon_1 y_1^2 - \kappa_1^2 \varepsilon_2 y_2^2 , 
\]

\[
B_1 = 2 \kappa_1 \kappa_2 x_0^2 {\cal C}_3 
\]

\noindent
and

\[
B_0 = [x_0 \kappa_1 \kappa_2]^2 .
\]

%In Fig. 2 source, images, critical curves, and caustics for this model  
\noindent
In Fig. 2 the source, images, critical curves, and caustics for this model are 
shown. The values of the model parameters are $ \sigma = \gamma = 0.05 $, 
$ \phi = 36^\circ $, $ \varepsilon = 0.2 $, $ \kappa_0 = 1 $, and 
$ x_0 = 0.5 $. The critical curves and caustics are again inclined 
($ \phi \ne 0 $). For this model, one gets up to five images. We have tried 
with other $ \beta $-values (with different values for the other parameters) 
to find more than five images, but all of our ESNIS models always produce no 
more than five images. 

%The isochrones are shown in Fig. 4. 

%\begin{figure*}%[bhpt]
\begin{figure}%[bhpt]
\begin{center}
\fbox{\includegraphics[width=8.2cm,height=8.2cm]{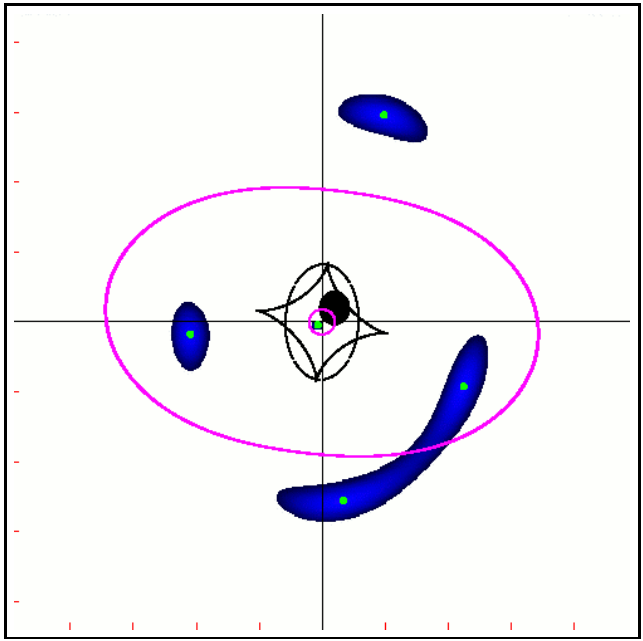}}
\caption{Source (large black dot near the centre), five images (one of them a 
long blue arc and one small image near the source), the critical 
curves (magenta ellipse and small circle), and the caustics 
(black ellipse and diamond) produced by a ESNIS model for following parameter 
values $ \beta = 1/2, \; \varepsilon = 0.2, \; \gamma = \sigma = 0.05, \; 
\phi = 36^{\circ}, \kappa_0 = 1 , \; x_0 = 1/2 $. 
The green dots are the positions of the images for a point source. 
The position of the extended source is $ (0.2, \; 0.2) $. 
The critical curves and caustics are inclined, because $ \phi \ne 0 $. 
The window size is $ 500 \times 500 $ pixel. 
%The horizontal axis represents 
%$ x_1 $ or $ y_1 $, and the vertical $ x_2 $ or $ y_2 $.
}
\end{center}
\end{figure}
%\end{figure*}

% XFGLenses:
% E = 2.5, theta = 0, alpha = 0.5, epsilon = 0.2, gamma = 0.2, sigma = 0.05
% phi = 36, kappa = 1, C = 0.5, X = 0.65, Y = 0.45, R = 0.2

%\begin{figure*}%[bhpt]
%\begin{figure}%[bhpt]
%\begin{center}
%\includegraphics[width=8.4cm,height=8.4cm]{Figure4.ps}
%\caption{Isocrones with images for the ESNIS model of Fig 3.
%The dots over the extended images are the images for a point source. 
%In this case there are 5 images: one little image close to the centre, 
%two in the arc, and two in the other extended images.} 
%\end{center}
%\end{figure}
%\end{figure*}

\noindent
We have designed XFGLENSES, an interactive program intended to visualize and 
model gravitational lenses \citep{frutos}. The modelling part of this program 
is not finished yet. The Figures 1, and 2 were generated with this software. 
The interested reader can download it from 

\begin{verbatim}
http://www.tat.physik.uni-tuebingen.de/~frutto/.
\end{verbatim}

\noindent
On this Website, there is also extensive information about the software.
To solve the GLEs, the Brent method was implemented in 
the program. To visualize the extended images, we use the Kayser-Schramm 
method \citep{schkay, kaysch}.

\section{Conclusions}

\noindent
Although the ESNIS model generated from elliptical potentials have 
an ellipticity restriction, these models are very useful in gravitational 
lens theory. The polynomial reduction presented here is useful not only to 
invert the GLE, but also to find the critical curves and caustics of a given 
ESNIS model. Moreover it could be useful to model gravitational lenses with 
small ellipticities. This class of models always produces up to five images. 
XFGLENSES can be used to visualize the results of this class of models. 
To see the capabilities and models that this software has available, visit 
the Website mentioned above.

% Sie stehen nämlich: 
% - direkt vor dem Vollverb =>  
% Sandra always, never, sometimes, usually, often 
% takes her dog for a walk after school
% - nach dem Hilfsverb und vor dem Vollverb => 
% Sandra doesn't usually take her dog for a walk in the evening. 
% Sandra can't always take her dog for a walk in the evening.
% Sandra must never take her dog for a walk. 
% nach am, is, are 
% she is always nice

\section{Acknowledgements}

\noindent
The author would like to thank the German Academic Exchange Service 
(DAAD: Deutscher Akademischer Austauschdienst) for its financial support and 
Professor Hans Ruder for the invitation to a research stay in T\"ubingen.

\bsp

\label{lastpage}
\end{document}